\documentstyle[sprocl,epsf]{article}

\def\mgut{M_{\rm{GUT}}}

\def\R{R_{b\tau}}
\def\yb{Y_b}
\def\ytau{Y_\tau}
\def\mz{M_Z}
\def\yt{Y_t}

\def\eea{\end{eqnarray}}
\newcommand{\p}{\partial}
\newcommand{\bra}{\langle}
\newcommand{\ket}{\rangle}

\newcommand{\tr}{\mbox{Tr}\,}
\newcommand{\nn}{\nonumber}

\begin{document}

\begin{flushright}
hep-ph/9705237
\end{flushright}

\vspace{5mm}

\title{Dynamical Gauge Bosons and Grand Unified Theory}
\author{MASAKO BANDO}
\address{Aichi University, Miyoshi, Aichi, 470-02, JAPAN \\
E-mail bando@aichi-u.ac.jp
}


\maketitle\abstracts{
\begin{enumerate}
\item  Introduction
\item  Dynamical Gauge Bosons and Compositeness Condition
\item  Is Standard Symmetry Asymptotically Free?
\item  Infrared Fixed Point Structure in ANF Theories
\item  Future Aspects
\end{enumerate}
My talk is based on the papers written with
J.~Sato and K.~Yoshioka~\cite{yosioka}, T.~Takeuchi,
T.~Onogi and J.~Sato.\cite{bost} and Y.~Taniguchi and
S.~Tanimura~\cite{tanitani}.
}

\section{Introduction}
The idea of dynamically generated gauge bosons  has
 long history since  1960's.
Originally Bjorken argued that gauge bosons are  NG bosons
responsible for spontaneous breaking of the Lorentz invariance.\cite{Bjorken}
Since then attempts to generate gauge bosons dynamically using
various models especially, NJL-type models were investigated
by many authors.
\cite{eguchi,terazawa,suzuki,georgi}.
However it was very hard to
get massless vector bound states and people had to be content
with  an approximate gauge symmetry
generation where the masses  of vector bound states are very small compared
with the relevant scale.
This is because
gauge symmetry has generically its profound background and is hardly
generated unless it exists from the first in the original system.
\cite{WeinbergWitten,KugoUehara}

The second approach to dynamical gauge bosons is the Berry phase mechanism,
which made clear that a gauge symmetry can be induced
when the system has two kinds
of degrees of freedom, fast and slow ones.
During evolution of slow variable,
fast one causes a static gauge field acting on slow one
\cite{Berry,TT,kikkawa,kikkawaTamura} and a gauge field is induced
by a mechanism of the Berry phase.
The Berry phase mechanism made clear that
the generation of gauge symmetry is a common phenomena which occurs
whenever we reduce the degrees of freedom of a system.
However, even in this case gauge symmetry is exact only when the
energy levels associated with the fast variables can be completely
neglected. This had been always a serious problem of any dynamical gauge boson
until  the notion of hidden local symmetry came into play.
The hidden local symmetry was first found in the context of
supergravity theory,
a l\'a the non-linear sigma model
\cite{cjulia,N=8} (for a review see Ref.18).
Thanks to the hidden gauge symmetry,
gauge invariance is guaranteed exactly and massless  vector bosons
are possible to be generated dynamically at quantum level and
they can be identified as corresponding gauge bosons.
As such an  example, we proposed the possibility that vector mesons
in QCD ($\rho$
or $A_1$ mesons) may be identified as Higgsed dynamical gauge bosons
of hidden local $SU(2)_V$ symmetry in the non-linear chiral
$SU(2)\times SU(2)/SU(2)_V$ Lagrangian.\cite{PhysRep,bky}
Thanks to the Recent arguments of duality, the idea
becomes now very popular
that gauge symmetries might not be fundamental
and they will appear as the consequences of
non-perturbative dynamical effects.
\cite{Seiberg}

With hidden local symmetry at hand,
attempts to generate gauge fields associated with hidden local
symmetry  were made using $CP^{n-1} $ models.
\cite{KTU}
and it was found that in  2- or 3-dimensional models,
gauge fields acquire their own kinetic terms via quantum effects,
and their  poles
are developed dynamically \cite{DVL,AA}.
However, these attempts have not been successful in four dimensions
; they can be generated only in cutoff theories.
\cite{KTU}
Here we would like to propose  that
cutoff theories are sufficient as
low energy effective theories:
they behave just like elementary gauge bosons in the
low energy region. The only difference of dynamical gauge bosons from
elementary ones  is  the
existence of the compositeness condition at some scale $Lambda$
below which they behave as if they are asymptotically non-free (ANF)  gauge
fields.
We shall explain this in the
section 2. Then we argue that ANF gauge theories may provide another
 possible scenario of the  ``standard
model''  below GUT scale (section 3), where we shall see that
infra-red stable points play essential roles (section 4).
Final section is devoted to future aspects and related problems.

\section{ Dynamical Gauge Bosons and Compositeness Condition}
Usually gauge interactions are required to be asymptotically free
(AF), otherwise the theory will become trivial
because of the existence of the Landau singularity at scale
$ \Lambda_L $.  We interpret this  singularity as an indication of
the so-called compositeness condition.
This idea of the compositeness condition at finite scale was first
argued by Bardeen, Hill and Lindner~\cite{BHL},
who claimed that the composite Higgs bosons is characterized
by the vanishing of the wave-function renormalization factor,
$ Z = 0 $, which is translated into the divergence of the rescaled Yukawa
coupling, $ y_r = \infty $ by rescaling the Higgs fields.
Similar arguments may be applied to dynamical gauge bosons and
the divergence of running gauge couplings at high energy scale $ \Lambda $
can be interpreted  as a
compositeness condition for gauge bosons.

Let us consider a simple model~\cite{tanitani}
in which dynamical gauge bosons
are generated associated with the hidden local symmetry in non-linear
$SU$($N_f$) / [$SU$($N_c$) $\times$ $SU$($N_f-N_c$)] Grassmannian like
model. Consider $ N_c \times N_f $ complex scalar fields $ \phi_{a i} $
and auxiliary $SU(N_c)$ fields $ A_{\mu} $ coupled to the
index $ a = 1 , \cdots , N_c $. The  matter fields $ \phi $
are transformed under
$ SU(N_f)_{\rm global} \times SU(N_c)_{\rm hidden\;local}$
as
$ \phi (x) \to h(x) \phi(x) g^\dagger $,
where
$ h(x) \in SU(N_c)_{\rm hidden\;local}$ and
$ g \in SU(N_f)_{\rm global}$~\cite{tanitani}.
The Lagrangian is given as
\begin{eqnarray}
     {\cal L} &=&
     \left( D_\mu \phi \right)^\dagger_{i a}
     \left( D^\mu \phi \right)_{a i}
     -
     \lambda_{a b}
     \left(
         \phi_{b i} \phi^\dagger_{i a}
         - \frac{N_f}{\omega} \delta_{a b}
     \right)
     + {\cal L}_{\rm GF + FP},
     \label{eqn:lagrangian}
\end{eqnarray}
with the covariant derivative $D_\mu \phi$;  $D_\mu \phi = \p_\mu \phi - i
A_\mu \phi$.
Here
$ \omega $ is a dimensionful coupling
and  ${\cal L}_{\rm GF + FP}$
 is the gauge fixing and the FP ghost term for $A_\mu$.
The hidden local gauge fields  $A_\mu$ do not have their kinetic terms
and represent redundant degrees of freedom of the model.
The $N_c \times N_c$ hermitian scalar field $\lambda_{a b}$ is the
Lagrange multiplier imposing the constraint;
$\phi_{a i} \phi^\dagger_{i b} = \frac{N_f}{ \omega } \delta_{a b}.
$. With this constraint $A_\mu$ can be eliminated by substituting the
equation of motion;
$ A_\mu = -\frac{i\omega}{2N_f}(\p_\mu \phi \, \phi^\dagger
-  \phi \, \p_\mu \phi^\dagger)$ and we obtain the following form:
\begin{eqnarray}
     {\cal L} &=&
     \tr \left[ \p_\mu \phi^\dagger
     \p^\mu \phi
     +
     \frac{\omega}{4N_f}
     (\phi^\dagger \p_\mu \phi-\p_\mu \phi^\dagger \,\phi)^2
     \right],
\end{eqnarray}
which is the original non-linear sigma model
without hidden local symmetry.

There are two alternatives in identifying the massless mode in $A_\mu$.
We can identify it as~\cite{Kugo-Townsend,KTU}; (1) Goldstone mode of
broken $SU(N_c)$ local symmetry, (2) Wigner mode of
$SU(N_c)$ gauge symmetry.
In the broken phase (1), the breaking of $SU$($N_c$) local
symmetry by the V.E.V. of
$\bra \phi \ket = \sqrt{N_f} v \left ( \delta_{a b} \>,\> 0\right)$
is accompanied by the breaking of $SU(N_f)$ symmetry.
On the other hand, in the symmetric phase (2), $A_\mu$
is a massless vector boson and no symmetry should be broken.
The phase of the model is determined by the effective potential.
Substituting the V.E.V.s of $\phi$ and
$ \lambda $, $\bra \phi \ket = \sqrt{N_f} \, v \, (\delta_{a b} , 0 )$,
$\bra \lambda \ket= \lambda \, \delta_{a b}$, into (\ref{eqn:lagrangian}),
 the effective potential is given in the $1/N_f$
approximation as
\begin{equation}
        V_{\rm eff}
        = N_f N_c
        \left[
                \lambda
                \left( v^2 - \frac{1}{ \omega } \right)
                + \int \frac{d^4 k}{i(2\pi)^4} \ln (\lambda-k^2)
        \right].
\end{equation}
This potential coincides with that of the $CP^{N-1}$ model~
\cite{PhysRep,KTU} within this approximation,
and the ground state is determined by the equations
\begin{eqnarray}
        &&
        \frac{1}{N_f N_c} \frac{\p V_{\rm eff}}{\p \lambda}
        =
        v^2 - \frac{1}{ \omega }
        + \int \frac{d^4 k}{i(2\pi)^4} \frac{1}{\lambda-k^2}
        = 0,
        \label{stationary}
        \\
        &&
        \frac{1}{N_f N_c} \frac{\p V_{\rm eff}}{\p v}
        =
        2\lambda v = 0.
\end{eqnarray}
with $ \Lambda $ being  a cutoff to define the integration.
Then we have,
\begin{eqnarray}
  v^2 - f ( \Lambda, \, \lambda)
  = \frac{1}{ \omega_{\rm r} ( \mu ) }
  - \frac{1}{ \omega_{\rm cr}(\Lambda , \, \mu )},
\label{cutoff}
\end{eqnarray}
where
\begin{eqnarray}
&&
\frac{1}{\omega_{\rm r}}
\equiv
\frac{1}{ \omega }
- \int \frac{d^4 k}{i(2\pi)^4} \frac{1}{\mu^2 - k^2},
\\
&&
\frac{1}{\omega_{\rm cr}}
\equiv
\int \frac{d^4 k}{i(2\pi)^4} \frac{1}{      - k^2}
- \int \frac{d^4 k}{i(2\pi)^4} \frac{1}{\mu^2 - k^2},
\\
&&
f(\Lambda , \, \lambda)
\equiv
\int \frac{d^4 k}{i(2\pi)^4} \frac{1}{        - k^2}
- \int \frac{d^4 k}{i(2\pi)^4} \frac{1}{\lambda - k^2}.
\end{eqnarray}
$ \omega_{\rm r} $ is a redefined coupling
at the renormalization point $ \mu $.
If the dimension is less than four,
$\omega_{\rm r}$, $ \omega_{\rm cr} $ and $ f $ remain finite
in the limit $ \Lambda \to \infty $~\cite{PhysRep,KTU}.
However, in four dimensions they  become divergent,
and so Eq.~(\ref{cutoff}) makes sense
only for the finite cutoff $ \Lambda $.
The function $ f $ is understood
as a non-negative function of $ \lambda $,
vanishing at $ \lambda = 0 $.
The critical coupling $ \omega_{\rm cr} $ which separates two phases;
(1) broken phase : $ \lambda = 0, \;\; v \neq 0$,
         when $ \omega_{\rm r} < \omega_{\rm cr} $,
 (2) symmetric phase :
        $ \lambda \neq 0, \;\; v = 0 $,
        when $ \omega_{\rm r} > \omega_{\rm cr} $.
The critical coupling $ \omega_{\rm cr} $
is proportional to the reciprocal of the logarithmic divergence:
$ \omega_{\rm cr}^{-1} \sim \log \Lambda $.
Thus $\omega_{\rm cr}$ becomes smaller as the cutoff $ \Lambda $
becomes larger and the symmetric phase is always realized by taking $ \Lambda $
to be sufficiently large.
In this phase the scalar field $\phi$ is massive
($m^2=\lambda$), whereas the vector $A_\mu$ becomes massless.
Hence this composite vector field is stable.
In the broken phase, the $\phi$ field becomes massless
(NG bosons associated with the broken symmetries),
while the gauge field $A_\mu$, if it exists, becomes massive
 through the Higgs mechanism.
This massive gauge boson is unstable as it can decay into two
$\phi$ bosons.

To see the generation of massless gauge bosons explicitly in
the symmetric phase with nonzero $\lambda$
we obtain the effective Lagrangian of the
composite gauge boson by calculating the one-loop Feynman diagrams
which contribute to the leading order terms of expansion with respect
to $N_c/N_f$;
\footnote{We use a dimensional regularization method
in order to preserve the hidden gauge invariance.
We then replace the regularization parameter $1/\bar{\epsilon}$ with
cutoff $\ln \Lambda^2$.}
\begin{eqnarray}
&&\ \ {\cal L}\ \ \ \ \ \  =\ \ \ \
 \tr \left( |\p_\mu \phi -i A_{\mu} \phi|^2
- \lambda |\phi|^2 \right)
-\frac{1}{4} Z_0 \left(\p_\mu A_\nu^A - \p_\nu A_\mu^A \right)^2 \nn\\
&&-\frac{1}{2} Z_1 f^{ABC} (\p_\mu A_\nu^A - \p_\nu A_\mu^A)
A_\mu^B A_\nu^C
-\frac{1}{4} Z_2 f^{EAB} f^{ECD} A_\mu^A A_\nu^B A_\mu^C
A_\nu^D\nn\\
\label{eqn:effective-lagrangian}
\end{eqnarray}
with $Z$ factors, $Z_0 = Z_1 = Z_2 = \frac{1}{16\pi^2}
\frac{1}{6} N_f \ln (\frac{\Lambda^2}{\mu^2})$, which
are dependent on
$\Lambda$. In conventional normalization,  $ Z(\mu)=1 $ at any scale
$ \mu $. Thus, if we redefine the field $A_{{\rm r}\,\mu}$ so as to
normalize the kinetic term in (\ref{eqn:effective-lagrangian}):
, $A_{{\rm r}\,\mu}= \sqrt{Z_0} A_\mu$,
the renormalized coupling is determined as
\begin{eqnarray}
g_{\rm r} &=& \frac{1}{\sqrt{Z_0}} = \frac{Z_1}{(\sqrt{Z_0})^3}
= \frac{\sqrt{Z_2}}{Z_0},
\label{eqn:renormalized_coupling}
\end{eqnarray}
which confirms the gauge invariance.
Note that the $Z$ factors vanish at $\mu = \Lambda$, at which
the dynamically generated kinetic as well as induced interaction terms
 disappear.
We call this the compositeness condition at the cutoff $\Lambda$.
This, in turn, indicates that the running coupling $g_r$
in (\ref{eqn:renormalized_coupling})
becomes infinity at scale $\Lambda$,
implying that the theory is an ANF theory~\cite{hasenfratz}.
The beta function obtained from (\ref{eqn:renormalized_coupling})
properly reflects this ANF behavior,
\begin{eqnarray}
\beta(g_{\rm r}) = \mu \frac{\p}{\p\mu} g_{\rm r}
= \frac{g_{\rm r}^3}{16\pi^2} \frac{1}{6}N_f,
\label{eqn:beta-fn}
\end{eqnarray}
from which we can see that the beta function (\ref{eqn:beta-fn})
of the composite theory coincides with that of the elementary theory
(the low energy theory with elementary gauge field),
\begin{eqnarray}
        &&
        {\cal L}
        =
        \tr \left(
          |D_\mu \phi|^2 - \lambda | \phi |^2
        \right)
        - \frac{1}{4}\tr (F_{\mu \nu})^2
        + {\cal L}_{\rm FP+GF},
        \label{eqn:elementary-lagrangian}
        \\
        &&
        \beta(g_{\rm r}) = \frac{g_{\rm r}^3}{16\pi^2}
        \left( \frac{1}{6}N_f - \frac{11}{3}N_c \right),
\label{eqn:coupling-el}
\end{eqnarray}
in the $N_c/N_f$ expansion.
With this approximation, the first term in (\ref{eqn:coupling-el})
dominates (the beta function is positive), suggesting ANF character.
The singularity at $\Lambda_L$ which appears in ANF gauge theory can be
interpreted as $Z=0$ in the composite theory, and above
the cutoff  $\Lambda_L$ the gauge field loses its identity
 as an elementary particle.
Thus the ANF gauge theory
(\ref{eqn:elementary-lagrangian}) is an indication of  a dynamically
generated gauge theory.

One may notice that a positive beta function in the above
always appears in the leading order of $ N_c / N_f $.
In this sense matters always have tendency to generate dynamical gauge
bosons while gauge bosons themselves make negative contributions
(the second term in (\ref{eqn:coupling-el}))~
 \cite{composite,composite2}. It should be noted that
if the beta function becomes negative due to the next order terms,
the theory does not satisfy the compositeness condition but is to be
understood as an elementary gauge theory without cutoff.
For this case, the $N_c/N_f$ expansion is no more applicable.

\section{Is Standard Symmetry Asymptotically Free?}
\begin{figure}[htbp]
  \begin{center}
    \epsfxsize=7.2cm \epsfysize=6.32cm \ \epsfbox{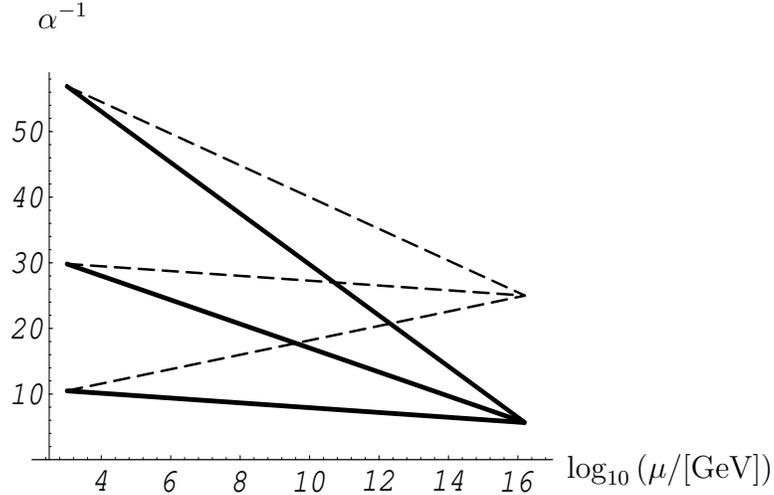}
    \put(-195,185){\large $\alpha^{-1}$}
    \put(5,15){\large $\log_{10} \left(\mu/[\mbox{GeV}]\right)$}
    \caption{Typical $\mu$ dependence of
      $\alpha_1(\mu)\,,\,\alpha_2(\mu)\,,\,\alpha_3(\mu)$ ~in the MSSM
      (dashed lines) and the MSSM + 1 EVF (solid lines).}
    \vspace*{-0.5cm}
    \label{fig:alpha}
  \end{center}
\end{figure}

In this section we discuss the possibility that the present standard
gauge group may not be AF.
The Minimal Supersymmetric Standard Model (MSSM) is very popular
because of its success in attaining gauge coupling
unification\cite{AMALDI:91}, which is crucial if one wishes
to construct a Grand Unified Theory (GUT).
Another attractive feature of the MSSM is the unification of
the $b$ and $\tau$ Yukawa couplings:
If one assumes at the GUT scale
\begin{equation}
\R(\mgut) = \yb(\mgut)/\ytau(\mgut) = 1,
\label{UNI1}
\end{equation}
the MSSM can reproduce
the experimental value of
$\R(\mz) \approx 1.8$ with an appropriate strength
of $ \yt(\mgut)$.
It is interesting
to investigate an extension of the
MSSM having ANF property.
In extending the MSSM by introducing extra matter superfields, we
must keep two things in mind:
(1) the matter superfields must be introduced in such
a way that gauge coupling unification
(and anomaly cancellation)~\cite{IBANEZ:81,HALL:91}
of the MSSM is preserved, and
(2) the fermion content must be compatible with the constraints
placed by LEP measurements, namely the so called Peskin--Takeuchi
constraint~\cite{PESKIN:90}.
The simplest way to satisfy these requirements is to
introduce 2 extra generations which form a generation--mirror
generation pair~\cite{MOROI:93}, which we will call
the 4th and anti--4th generations with
$SU(2)_L\times U(1)_Y$ invariant Dirac masses
{}~\cite{MAEKAWA:95},which we call Extended Supersymmetric Standard
Model (ESSM)~\cite{babupati:96}.
The typical behaviors of the running
gauge couplings are shown in Fig.\ref{fig:alpha}.
They are unified at the same scale in both cases but with
different unified couplings.

Then the question arises: Are the above predictions still preserve in
ESSM?
One immediate consequence of ESSM is that all three gauge couplings of
 $SU(3)_C, SU(2)_L$ and $ U(1)_Y$,  are ANF:
they become larger as they are evolved up to
coincide at the unification scale~\cite{THEISEN:88}.
We extend  the analysis of ESSM and study
how the existence of the extra generations will affect the
running of the Yukawa couplings of the 3rd generation fermions.
In addition to the $SU(2)_L \times U(1)_Y$ invariant masses,
we also couple the 4th and $\bar{4}$th generation fermions to the
two Higgs doublets in the same way as the other generations;
\begin{equation}
Y_U = Y_t,  \quad  Y_D = Y_b,  \quad Y_E = Y_{\tau},
 \qquad  Y_{\bar D} = Y_{\bar U} =  Y_{\bar E} = 0,
\label{yukawas}
\end{equation}
and set all the 1st and 2nd generation Yukawa couplings to zero.
As in the MSSM case, we will impose a unification condition on the
Yukawa couplings at $\mgut$ and determine the parameter range in which
our model can predict the correct top, bottom and $\tau$-lepton masses.
At this point one may think that such a program is doomed
to failure from the beginning.
Since the QCD coupling is ANF, the QCD enhancement
of $\R$ from $\mgut$ to $\mz$ will be even larger than the MSSM case
making it impossible to bring $\R(\mz)$ down to $\sim 1.8$
even with large Yukawa couplings.
However,  an $SO(10)$-GUT with an 126-Higgs predicts~\cite{CHANOWITZ:77}
\begin{equation}
\yt(\mgut) = \yb(\mgut) = \frac{1}{3}\ytau(\mgut) \quad   \rightarrow
 \quad \R(\mgut) = \frac{1}{3},
\label{UNI3}
\end{equation}
which is the unification condition we will adopt here.\footnote{%
An 126-Higgs is necessary to give a direct
Majorana mass term to the right-handed neutrino.}
In this case, the extra enhancement from
QCD is actually welcome since $\R$ must be enhanced by a factor of
$5 \sim 6$ to reproduce the experimental value of $\R(\mz)$.

The number of adjustable parameters is four:
the unification scale $\mgut$,
the unified gauge coupling $\alpha_{\rm{GUT}}$,
the unified Yukawa coupling $Y_{{\rm GUT}}$,
and the mixing angle of the low lying Higgs fields
$\tan\beta = v_2/v_1$ ($\sqrt{v_1^2 + v_2^2}=246\,{\rm GeV}$).
We restrict $\alpha_{\rm{GUT}}$ and $Y_{{\rm GUT}}$ to the region,
$\alpha_{\rm{GUT}} < 1.0, Y_{\rm{GUT}} < 0.7$, which
guarantees that the gauge and Yukawa couplings are kept still within their
perturbative regions throughout the evolution from $\mgut$ to $\mz$.
We will use the $\tau$-lepton mass to fix $\tan\beta$ from
In view of the relatively large coupling strengths near the
unification scale due to the ANFness,
we use the fully coupled
2-loop RGE's to  evolve the gauge and Yukawa
couplings. Taking the masses at $M_{EVF} =M_{SUSY} = 1\,\rm{TeV}$,
and by fixing  the values of $\alpha_{GUT}$ and $M_{GUT}$ in the
range allowed by gauge coupling unification, we calculate the evolution of
the Yukawa couplings for different values of $Y_{GUT}$.

\begin{figure}[tp]
\centering
\unitlength=0.8cm
\begin{picture}(14,9)
\unitlength=1mm
\put(4,78){$\R(\mz)$}
\put(110,30){$\log_{10}M_{GUT}(\rm{GeV})$}
\epsfxsize=12cm
\ \epsfbox{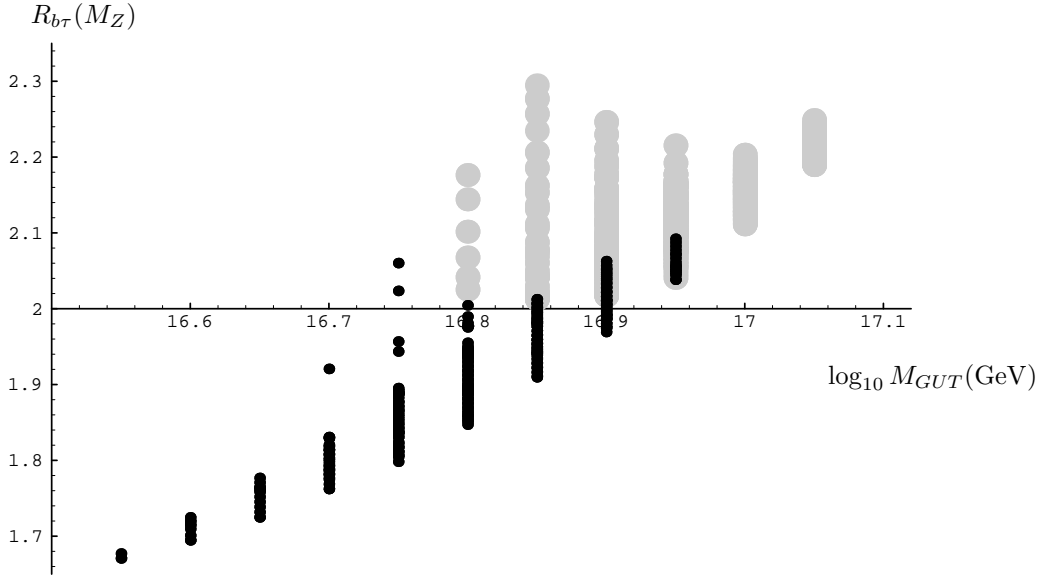}
\end{picture}
\vspace{-5pt}
\caption[]{
The dependence of $\R(\mz)$ on $M_{GUT}$ and $\alpha_{GUT}$.
The small-black, and large-gray circles indicate the ranges
$0.3 < \alpha_{GUT} \le 0.6$, and
$0.6 <  \alpha_{GUT} < 1$, respectively.
The dependence of $\R(\mz)$ on $Y_{GUT}$ is negligible.
}
\label{fig:Rbtau}
\end{figure}

The experimentally determined $\bar{\rm{MS}}$ running masses
of the $\tau$--lepton and the $b$ quark at $\mu = \mz$ are given by
{}~\cite{PDG:96},
\begin{eqnarray}
 &&m_{\tau}(M_{SUSY})= \frac{v}{\sqrt{2}}Y_{\tau}(M_{SUSY})\cos\beta
\nn  \\
&&m_{\tau}(\mz)= 1.75 \pm 0.01 \,\rm{GeV},\ \ \
m_b(\mz) =  3.1  \pm 0.4  \,\rm{GeV},
\label{eq:}
\end{eqnarray}
from which we conclude $\R(\mz) = 1.6 \sim 2.0$.
The calculated result of  $\R(\mz)$ within reasonable values of
 $\alpha_{GUT}$ and
$M_{GUT}$ is
shown in Fig.~\ref{fig:Rbtau}.
 From the figure we see that There is a large set of
$(\alpha_{GUT}, M_{GUT})$ values which keeps $\R(\mz)$ below 2.

We would like to stress that  the ANFness of the
gauge couplings has a strong focusing effect on the top (and also bottom)
Yukawa couplings
as they evolve down in scale and as a result,
the value converges to IR fixed point
by the time they reach the SUSY breaking scale $M_{SUSY}=
1\,\rm{TeV}$.
In the next section we shall show the explicit picture of the
 evolutions of the  $t$ Yukawa
couplings. Also we shall see that for  the case of the $\tau$ Yukawa coupling,
the situation is rather
different. Near the GUT scale it tends to focus itself
due to its larger size at $M_{GUT}$ (Recall $Y_{\tau}(M_{GUT}) = 3Y_{GUT}$.).
However, unlike the top and bottom Yukawa couplings,
once $Y_{\tau}$ becomes small at lower scales, it runs
slowly and does not quite converge to its IR fixed point ($y_{\tau}=0$).

\section{ Infrared Fixed Point Structure in ANF Theories}
We have shown  a possible
scenario of the standard gauge symmetry with ANF character and
mentioned that due to the ANF gauge
couplings the top Yukawa coupling is determined almost insensitive
to their initial values fixed at GUT scale $M_G$.
We would like to stress  that such  strong convergence of Yukawa
couplings to their infrared fixed points is a common feature appearing
in ANF theories.
This section is devoted to present how strongly the couplings are focused
into their infrared points in ANF theories and demonstrate the
structure of the renormalization group flow.  As illustrations we take the
supersymmetric standard models with AF and ANF gauge couplings and
compare them by concentrating on their infrared structure.

\begin{figure}[htbp]
  \begin{center}
    \epsfxsize=7.2cm \epsfysize=6.32cm \ \epsfbox{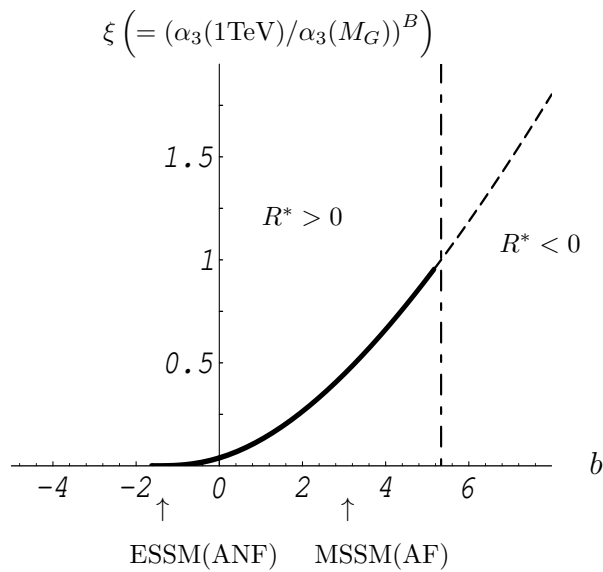}
 \put(-170,180){$\xi
      \left( = \left(\alpha_3(1 {\rm TeV})/\alpha_3(M_G)\right)^B\right)$}
    \put(-20,100){$R^* < 0$}
    \put(-110,110){$R^* > 0$}
    \put(14,16){\large $b$}
    \put(-150,0){$\uparrow$}
    \put(-80,0){$\uparrow$}
    \put(-160,-18){ESSM(ANF)}
    \put(-90,-18){MSSM(AF)}
    \caption{Typical behavior of suppression factor $\xi$ by taking
      $\alpha = \alpha_3$ }
    \vspace*{2mm}
    \hspace*{-1cm}$( \alpha_3(M_Z) = 0.12\,,\,c = \frac{16}{3} )$
    \label{fig:sup}
  \end{center}
\end{figure}

Before this let us consider a simple gauged Yukawa system with
one gauge coupling
$g$ and one Yukawa coupling $y$, whose 1-loop $\beta$-functions are as
follows:
\begin{equation}
 \frac{d \alpha}{d t} \!\! = \!\! - \frac{b}{2\pi} \alpha^2,\quad
  \frac{d \alpha_y}{d t}\!\! = \!\! \frac{\alpha_y}{2\pi}\, \Bigl( a
  \alpha_y - c \,\alpha \Bigr),
\end{equation}
where,
\begin{eqnarray}
  \alpha \equiv \frac{g^2}{4\pi} \,,&& \alpha_y \equiv
  \frac{y^2}{4\pi}\,,\qquad t = \ln \left(\frac{\mu}{\mu_0}\right) .
\end{eqnarray}
The system is AF (ANF) for
$b > 0  \;( b < 0 ) $ and always $a > 0\,,\,c \geq 0 $.
Then,
\begin{eqnarray}
  &&\frac{d R}{d t} = \frac{a}{2 \pi}\alpha R \left( R - R^* \right),
  \qquad  \left( R \equiv \frac{\alpha_y}{\alpha},\quad
   R^* = \frac{c-b}{a} \right).
  \label{eqn:R}
\end{eqnarray}
If we take $\,\mu_0 = M_G$ and
$\alpha(0) = \alpha(M_G)$, we get from (\ref{eqn:R})~\cite{bost}
\begin{eqnarray}
  \frac{R(t) - R^*}{R(t)} \!\! = \!\!
  \left(\frac{\alpha(t)}{\alpha(M_G)}\right)^B \left( \frac{R(M_G)
      - R^*}{R(M_G)}\right),\,\quad  B \equiv 1 - \frac{c}{b}.
      \label{eqn:Rsol}
\end{eqnarray}
where $R^*$ is an infrared fixed point if $R^*>0$ and
we see that the suppression factor
$\xi \equiv \left(\frac{\alpha (t)}{\alpha(M_G)} \right)^B$
, gives the criterion
on how fast $R$ approaches to $R^*$, depends only on the gauge
coupling~\cite{Ross}.
The $b$-dependence on the suppression factor $\xi$ is shown in
Fig.\ref{fig:sup},
from which we find a big difference between  AF ($b>0$) and ANF
($b<0$) cases. In the AF case the point $b=c$ above which $B$ becomes
negative and  $\xi$ becomes larger than 1 and $R^*(<0)$ is no more an
infrared fixed point~\cite{kugo}. On the other hand,
in the ANF case there is always
a nontrivial infrared fixed point $R^* (>0)$ and the convergence to
$R^*$ becomes much better.
Let us compare the case of MSSM(AF) with that of the ESSM(ANF)~
 \cite{MPP,MOROI:93,bost}.

Let us compare the typical values corresponding to  AF and ANF cases,
respectively, by taking the realistic values of $\alpha =
\alpha_3\,,\,\alpha_y = \alpha_t$;
\   1) \  MSSM: $b = 3$, $ c = \frac{16}{3}$ $\Rightarrow$
$\xi  \sim 0.423$  and
 2) \ ESSM: $b =-1$, $c = \frac{16}{3}$ $\Rightarrow$
$\xi \sim 10^{-6}$.

\begin{figure}[htbp]
  \begin{center}
    \epsfxsize=7,2cm \epsfysize=6,32cm \ \epsfbox{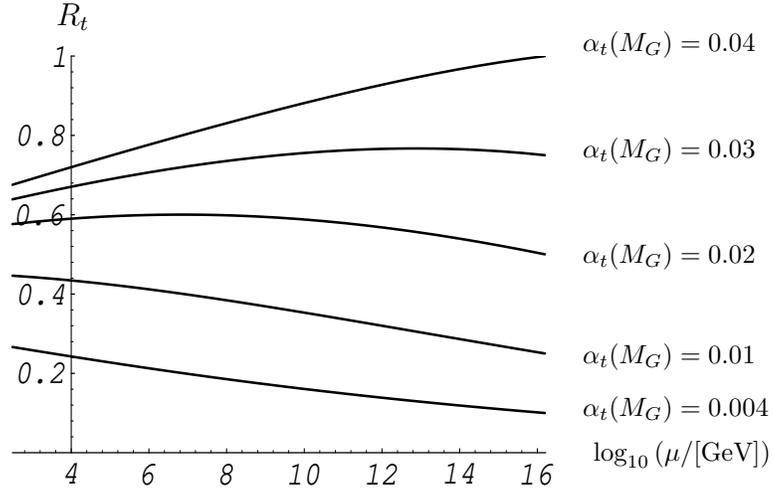}
    \put(15,15){{$\log_{10} \left(\mu/[\mbox{GeV}]\right)$}}
    \put(-188,180){{\large $R_t$}}
    \put(11,170){$\alpha_t (M_G) = 0.04$}
    \put(11,130){$\alpha_t (M_G) = 0.03$}
    \put(11,90){$\alpha_t (M_G) = 0.02$}
    \put(11,52){$\alpha_t (M_G) = 0.01$}
    \put(11,32){$\alpha_t (M_G) = 0.004$}
    \caption{$R_t$ ~in the MSSM ~
      ($M_G = 1.6\times 10^{16}\;\mbox{GeV}\,,\; \alpha_{GUT} = 0.04$)}
 \label{fig:t/3-MSSM}
  \end{center}
\end{figure}

\begin{figure}[htbp]
  \begin{center}
    \epsfxsize=7.2cm \epsfysize=6.32cm \ \epsfbox{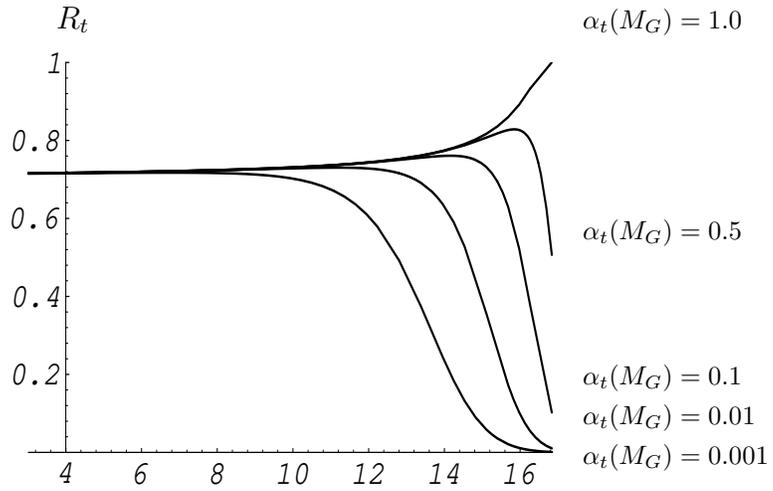}
    \put(-188,180){{\large $R_t$}}
    \put(11,180){$\alpha_t (M_G) = 1.0$}
    \put(11,100){$\alpha_t (M_G) = 0.5$}
    \put(11,45){$\alpha_t (M_G) = 0.1$}
    \put(11,30){$\alpha_t (M_G) = 0.01$}
    \put(11,15){$\alpha_t (M_G) = 0.001$}
    \caption{$R_t$ ~in the MSSM
      + 1 EVF ~($M_G = 7.0\times 10^{16}\;\mbox{GeV}\,,\; \alpha_{GUT}
      = 1.0$)}
    \label{fig:t/3-ANF}
  \end{center}
\end{figure}
The suppression factors can be read off from Fig.\ref{fig:sup}
(indicated by arrows). We can see more clearly the situation by
comparing the $\mu$-dependence of $\alpha_t/\alpha_3$ in AF and ANF
cases. In the ESSM case the convergence to the infrared fixed
point is much better than that in the MSSM  (Fig.\ref{fig:t/3-MSSM})
and its fixed point value
depends very weakly on the initial value at $M_{GUT}$ (Fig.\ref{fig:t/3-ANF}).
This is because of the effect from the gauge couplings which becomes
very large at $M_G$.

The RG flow in $(\alpha_3,\alpha_t)$ plane of those 2 cases
behaves quite differently each other, where the ANF case is
in remarkable contrast to the AF
case, the details of which is left to the reference~\cite{yosioka}.
In the case of ANF, the ratio  $\frac{\alpha_t}{\alpha_3}$ evidently has
infrared fixed point in all the region in the
$(\alpha_3,\alpha_t)$ plane and one loop approximation becomes more
and more available in the infrared region.
This is in remarkable contrast to the AF case.
This kind of infrared stability is very attractive and
may give us a feasibility of determining low-energy Yukawa coupling constants.
Finally  we make a comment on the infrared structure of $R_\tau$.
$R_\tau$ does not reach to the non-trivial infrared fixed point
but to zero in the low energy limit.
On the other hand, the top Yukawa coupling constant is
quite remarkable ( Fig.\ref{fig:t/3-ANF}); we find that
it indeed reaches  to their fixed point, whose value
 hardly be affected by the initial values at $M_G$
because of the ANF gauge couplings.  We are sure
that the infrared fixed points obtained from these solutions are
physically significant and provide us with reliable low-energy
parameters. In the case at hand, by using these fixed point solutions and the
experimental value of $\alpha_3(1 \mbox{TeV}) \sim \, 0.093$ ,
we get for example \footnote{The value $\tan \beta \sim 58$ is
determined from the experimental value $m_\tau(M_Z)$ \cite{bost}. },

\begin{eqnarray}
  m_t(M_Z) \sim 178 \,\mbox{GeV}\,,&& m_b(M_Z) \sim 3.2 \,\mbox{GeV}
  \\[3mm] &&\hspace*{-2cm}( \,\tan \beta \sim 58\, )
  \nonumber
\end{eqnarray}

\noindent These values are certainly consistent with the experimental
values~\cite{PDG:96},
\begin{eqnarray}
m_t(M_Z) \sim 180 \pm 10 \,\mbox{GeV}\, , m_b(M_Z) \sim 3.1 \pm 0.4
\,\mbox{GeV}\ .
\end{eqnarray}

\section{ Future Aspects }
The points in the parameter-space of
Lagrangian, to which the renormalization flow of the parameters
in the Lagrangian converge in the high (low energy) limit,
are called ultraviolet (infrared) fixed points and they play
important roles in
characterizing the physics.
The theory is  called ``asymptotically free''
(AF)  when its ultraviolet fixed points of the parameters
are zero  (trivial ultraviolet fixed points).
If we trace back to the low energy region of AF theory, on the other
hand, naive continuation  leads larger
and larger couplings at lower energy, showing
infrared divergence. In contrast to AF theories, we have seen that
ANF theories have nontrivial infrared fixed points which have
close relations to low-energy physics.
Usually ANF theories have been considered to be
unphysical and
people thought that
couplings should never diverge in high energy limit.
However we have demonstrated that ANF theories may provide possible candidates
 for models to unify all existing particles and fields.
This may be guaranteed so far as the ultraviolet cutoff
$\Lambda$ has a well defined physical meaning  and if (
at least) one of the couplings becomes infinity at a certain scale
$\Lambda$, which is just an indication of some new physics above $\Lambda$.
Recently in some theories having nontrivial  fixed points,
it is found that
there exist some relations between different kinds of coupling
constants, and the symmetry structure of the system is enlarged
in this limit. In this context we may hope that infrared structure
appearing in ANF theories may provides important information on the
low energy physics.

Then we can proceed further to consider the scenario
of GUT with ANF standard gauge groups.
As an example, we have seen that a vector-like family pair
added to MSSM attains gauge unification at 2-loop level and
reproduces experimental bottom tau ratio quite easily, and
further it connects the top quark mass to an infrared
fixed point.
The existence of extra fermions have been discussed from various points
of view; in deriving CP violation, dynamical SUSY breaking,
hierarchical mass matrix,
especially we are aware that any GUT model motivated by
string models or supergravity models, predicts additional fermions
quite naturally.
At present it is very difficult to predict their masses,
but we must not exclude the cases in which they are even
below 1 TeV~\cite{fujikawa,inoue}.
If it is so, we may find their consequences
in near future. ANF model predicts that the strong coupling constant
decreases very slowly or even increases at higher scale.
This will be tested by measuring
the excess of jet production cross section
around 1 TeV or so in the future supercollider experiments.

We have argued that any non-linear sigma model has its own hidden
local symmetry and there may be a possibility that
the gauge bosons associated with this hidden local
symmetry are generated dynamically.
However this is not enough for those vector fields to be real gauge
bosons. If one introduces matter fields, for example, there is no reason that
those matters couple to those gauge bosons with universal strength and
they couple to hidden gauge bosons
with arbitrary strength.
The only case for the hidden gauge bosons
to have universal couplings, seems to assign all
the existing matters as NG bosons living in the coset space G/H.
Then the universality of gauge  couplings  is always guaranteed.
This possibility is very attractive and once investigated
elaborately in
1980~\cite{N=8} within the framework of $N=8$
 supergravity theories, where the hidden local symmetry has
played an important role.
In this scenario it is assumed that
all the existing matters  and fields except graviton are bound
states composed of $N=8$ gravity multiplets~\cite{bandosatouehara,N=8}.
Of course there are many problems; anomaly, chirality, the cosmological
constant, etc. I do hope that we shall be able to overcome for all
those proble in the near
future.

\section*{Acknowledgments}
I would like to thank T.~Kugo, T.~Yanagida  and the collaborators,
J.~Sato, Y.~Taniguchi, K.~Yoshioka, S.~Tanimura, T.~Onogi and
T.~Takeuchi  for their  valuable comments and discussions.
Also the conversations with  T.~Takahashi, N.~Maekawa, H.~Haba
 were helpful and encouraging.
The main part of this work was done during the Kashikojima Workshop
held in August 1996 at Kashikojima Center,
which was supported by the Grand-in Aid for Scientific Research
(No. 07304029) from the Ministry of Education, Science and Culture.
I am supported in part by the Grand-in Aid for
Scientific Research
(No. 06640416) from the Ministry of Education,
Science and Culture.

\makeatletter
\def\@JLone<#1,#2>{#1}
\def\@JLtwo<#1,#2,#3>{#2}
\def\@JLyear<#1,#2,#3,#4>{#3}
\def\@JLpage<#1,#2,#3,#4>{#4}
\def\JL#1{{\em \@JLone<#1>}\ {\bf \@JLtwo<#1>}, \@JLpage<#1> (\@JLyear<#1>)}
\def\@Jpage<#1,#2,#3>{#3}
\def\andvol#1{{\bf \@JLone<#1>}, \@Jpage<#1> (\@JLtwo<#1>)}
\def\PTP#1{Prog.\ Theor.\ Phys.\ \andvol{#1}}
\def\JPSJ#1{J.~Phys.\ Soc.\ Jpn.\ \andvol{#1}}
\def\PR#1{Phys.\ Rev.\ \andvol{#1}}
\def\PRL#1{Phys.\ Rev.\ Lett.\ \andvol{#1}}
\def\PL#1{Phys.\ Lett.\ \andvol{#1}}
\def\NP#1{Nucl.\ Phys.\ \andvol{#1}}
\def\JMP#1{J.~Math.\ Phys.\ \andvol{#1}}
\def\IJMP#1{Int.\ J.~Mod.\ Phys.\ \andvol{#1}}
\def\CMP#1{Commun.\ Math.\ Phys.\ \andvol{#1}}
\def\JP#1{J.~of Phys.\ \andvol{#1}}
\def\ANN#1{Ann.\ of Phys.\ \andvol{#1}}
\def\NC#1{Nouvo Cim.\ \andvol{#1}}
\def\PREP#1{Phys, Report\ \andvol{#1}}
\def\ZP#1{Z.~Phys.\ \andvol{#1}}
\makeatother

\end{document}